\documentclass{pasj00}
\draft

\begin{document}
\SetRunningHead{Honda et al.}{Enrichment of Heavy Elements in S~15--19 in the Sextans Dwarf Spheroidal Galaxy}
\Received{2010/12/02}
\Accepted{2010/12/20}
%\Published{}%{yyyy/mm/dd}

\title{Enrichment of Heavy Elements in the red giant S~15--19 in the Sextans Dwarf Spheroidal Galaxy$^*$}

%%% begin:list of authors
% Do NOT capitalize all letters in "textsc".
\author{Satoshi \textsc{Honda} %
%  \thanks{Example: Present Address is Kwasan Obs.}}
}
\affil{Kwasan Observatory, Kyoto University, Ohmine-cho Kita Kazan, Yamashina-ku, Kyoto 607-8471}
\email{honda@kwasan.kyoto-u.ac.jp}

\author{Wako \textsc{Aoki}, Nobuo \textsc{Arimoto}}
\affil{National Astronomical Observatory of Japan, 2-21-1 Osawa, Mitaka, Tokyo 181-8588}
\affil{Department of Astronomical Science, The Graduate University of Advanced
Studies, Mitaka, Tokyo 181-8588}
\email{aoki.wako@nao.ac.jp, arimoto.n@nao.ac.jp}

\and
\author{Kozo {\sc Sadakane}}
\affil{Astronomical Institute, Osaka Kyoiku University, 4-698-1 Asahigaoka, Kashiwara-shi, Osaka 582-8582}\email{sadakane@cc.osaka-kyoiku.ac.jp}
%%% end:list of authors

%%% Please use the following style in case that sorting by 
%%% affilation is impossible. 
%
% \author{%
%   D-Firstname \textsc{D-Familyname}\altaffilmark{1}
%   E-Firstname \textsc{E-Familyname}\altaffilmark{1,2}
%   and
%   F-Firstname \textsc{F-Familyname}\altaffilmark{2}}
% \altaffiltext{1}{Address of Institute}
% \email{ddddd@xxx.xxx.xx.xx}
% \email{eeeee@xxx.xxx.xx.xx}
% \altaffiltext{2}{Address of Institute}

%% `\KeyWords{}' always has to be placed before `\maketitle'.
 %Do NOT move this preamble from here!

\maketitle

\begin{abstract}
  We determined chemical abundances of the Extremely Metal-Poor (EMP)
  star S~15--19 ([Fe/H]$=-3.0$) in the Sextans dwarf galaxy. While
  heavy neutron-capture elements (e.g., Ba) are generally deficient in EMP stars in dwarf
  galaxies, this object was shown to have an exceptional over-abundance of Ba ([Ba/Fe]$\sim +0.5$) by a previous study, which
  is similar to those of r-process-enhanced stars found in the field
  halo. Our new high-resolution spectroscopy for this object for the
  blue region, however, reveals that no clear excess of r-process
  elements, like Eu, appears in this object. Moreover, a significant excess
  of carbon ([C/Fe]$= +1.0$) and a deficiency of Sr ([Sr/Fe]$ = -1.4$)
  are found for this object. Taking the variation of radial velocities 
  measured at the two different epochs into consideration, the origin 
  of the excesses of heavy neutron-capture elements in S~15--19 is not 
  the r-process, but is the s-process in an asymptotic giant branch 
  (AGB) star that was the binary companion (primary) of this object.
  Carbon- and s-process-enhanced material should have been transferred to the 
  surface of S~15--19 across the binary system. These results are 
  compared with carbon-enhanced metal-poor stars in the field halo.

\end{abstract}
\footnotetext[*]{Based on data collected at Subaru Telescope,
which is operated by the National Astronomical Observatory of Japan.}

\section{Introduction}

Dwarf galaxies around the Milky Way (MW) are believed to provide a
key to understanding the formation processes of the MW halo structure
and chemical evolution in small galaxies (e.g., Mateo et al. 1998;
Tolstoy et al.  2009). In order to determine the chemical composition
of individual stars of dwarf galaxies, high-resolution spectroscopy has been obtained for red giants in some nearby dwarf spheroidals with 8-10m class 
telescopes in the past decade.

While a large sample of red giants have been studied for some dwarf
spheroidal galaxies (e.g., Fornax: Letarte et al. 2010), abundance
measurements of extremely metal-poor (EMP) stars ([Fe/H]$\lesssim -3$)
are still quite limited. A few EMP stars have been studied by
\citet{shetrone01}, \citet{fulbright04}, and \citet{sadakane04}.
Recently, the chemical compositions of EMP stars in the Sextans dwarf
galaxy found by \citet{helmi06} were studied in detail by \citet{aoki09}. 
The sample of EMP stars in dwarf galaxies studied by high-resolution spectroscopy 
has been increasing owing to recent work \citet{cohen09} and \citet{cohen10} 
for Draco and Ursa Minor. An EMP star with
[Fe/H]$=-3.7$ was discovered in the Sculptor dwarf galaxy by \citet{frebel10a}. 
Moreover, EMP stars have been discovered in the
so-called Ultra Faint dwarf Galaxies (UFdG), for which abundance
measurements with high-resolution spectroscopy have been made only
quite recently (see below).

These studies suggest that the abundance ratios of $\alpha$-elements
to Fe (e.g., Mg/Fe) of EMP stars in most dwarf galaxies are as high as
those of halo field stars, while the ratios are sometimes
significantly lower at higher metallicity. However, low $\alpha$/Fe
ratios are at least found in EMP stars of Sextans \citep{aoki09}, and
further studies for a larger sample are required to derive any
conclusion on the origins of the chemical composition of these EMP stars.

Another interesting feature found by previous studies for very
metal-poor stars ([Fe/H]$\lesssim -2.5$) in dwarf galaxies is the low
abundances of heavy neutron-capture elements. While some stars with
overabundances of Ba ([Ba/Fe] $> 0$) were found by \citet{shetrone01} 
for [Fe/H]$\gtrsim -2$, no such object with lower metallicity exists in their
sample. \citet{fulbright04} and \citet{sadakane04} reported extremely
metal-poor ([Fe/H]$\sim -3$) stars with strikingly low Ba abundances
([Ba/Fe]$<-2.6$ for Draco 119, and [Ba/Fe]$=-1.3$ for Ursa Minor
COS~4, respectively). Low Ba (and Sr) abundances have also been found by
more recent studies of dwarf galaxy stars \citep{aoki09, cohen10} as
well as for UFdGs \citep{frebel10b, simon10}.

This makes a clear contrast to the neutron-capture elements in MW EMP
stars, in which a very large dispersion of the abundances of heavy
elements with respect to Fe is seen (e.g., McWilliam et al. 1995;
Honda et al. 2004; Fran{\c c}ois et al. 2007).  The abundance patterns
of heavy elements in EMP stars in the field halo that show over-abundances of
neutron-capture elements are usually similar to the solar-system
r-process pattern (e.g., Sneden et al. 2003).  This is interpreted as being 
a result of the short time-scale of the enrichment by the r-process events compared 
to s-process enrichment by evolved intermediate-mass stars
(AGB stars), although the astrophysical site of the r-process is still
not well understood. On the other hand, a large fraction of
carbon-enhanced metal-poor (CEMP) stars also show an enhancement of
neutron-capture elements \citep{aoki07}.  The origin of heavy elements
of these objects is the s-process in metal-poor AGB stars, even at low
metallicity. Such objects are interpreted as being a result of mass transfer
from a companion that was previously an AGB star, and has been provided with
carbon-rich material with a large excesses of heavy neutron-capture
elements produced by the s-process during AGB evolution \citep{busso99}.

Among the EMP stars in dwarf galaxies studied so far, the red giant
S~15--19 in Sextans has an exceptionally high Ba abundance
([Ba/Fe]$=+0.5$: Aoki et al. 2009). The value is as high as those of
the r-process-enhanced stars in the field halo (r-II stars: Beers \&
Christlieb 2005). If the origin of the Ba excess in S~15--19 is the
r-process, this indicates that a large scatter of
neutron-capture elements also appears in dwarf galaxies, as found in
the field halo.  However, no constraint on the origin of heavy
elements of this object was obtained by a previous study
\citep{aoki09} because no other neutron-capture element was measured
due to the limited wavelength coverage and data quality.

In order to determine the origins of heavy elements in S~15--19 in
Sextans, we obtained a new blue spectrum with higher quality, and
conducted detailed chemical abundance measurements.  This paper
reports on the result of our abundance study, and discusses the enrichment
processes of this object.

\section{Observations and Measurements}

High resolution spectroscopy was carried out for S~15--19 on 2010 Feb. 9
using Subaru/HDS \citep{noguchi02}.  The spectrum covers
3760-5490~{\AA} with a resolving power of 40,000 by 2$\times$2 on-chip
binning.  This spectral range includes atomic lines of Eu, Sr, and CH
molecular bands, which were not covered by a previous study 
(4400-7100~{\AA}: Aoki et al. 2009).  The echelle data were processed using the IRAF
software echelle package\footnote{IRAF is distributed by the National
  Optical Astronomy Observatories, which is operated by the
  Association of Universities for Research in Astronomy, Inc. under
  cooperative agreement with the National Science Foundation.}  in a
standard manner.  
The long exposure (8 hr) collected 980 photons per
1.8~km~s$^{-1}$ pixel at 4500~{\AA}. The spectrum is, however,
significantly affected by the sky background. The sky spectrum was
extracted from the region around the object on the slit, and subtracted. 
The signal-to-noise (S/N) ratio of the spectrum after sky
subtraction was 25 (per resolution element) at 4500~{\AA}.

Equivalent widths were measured by fitting Gaussian profiles.  The derived
equivalent widths are given in table 1.  Most of the line data were
taken from the list of \citet{aoki09} and \citet{aoki05}.  The
equivalent widths are compared with the previous measurements of Fe
{\small I} lines \citep{aoki09} for the overlapping wavelength range
(4400 -- 5490~{\AA}) in figure 1.  The results of the present work are slightly
larger (by 6\%) than those of \citet{aoki09}. However, the
discrepancy is smaller than the measurement errors (13 ~m{\AA} for a
S/N$\sim 25$ spectrum), indicating no evident systematic difference in the
equivalent widths between the two studies.

A measurement of radial velocity was made for selected clean iron
lines used in the equivalent-width measurements.  
The derived radial velocity is 223.60 $\pm$ 0.23 km s$^{-1}$.  
This value is by 3~km~s$^{-1}$ different from that obtained by a previous
study (226.05 $\pm$ 0.11 km s$^{-1}$: Aoki et al. 2009), suggesting that
S~15--19 belongs to a binary system.

\section{Abundance Analysis and Results}

Chemical abundance analyses were performed using the analysis program
SPTOOL developed by Y.Takeda (Takeda 2005 private communication),
based on Kurucz's ATLAS9/WIDTH9 (Kurucz 1993).  We computed spectra and
equivalent widths based on model atmospheres under the assumption of
local thermodynamic equilibrium (LTE).

We adopted the model atmosphere parameters (effective temperature,
gravity, microturbulent velocity, and metallicity) derived by a
previous study \citep{aoki09}, which are $T_{\rm eff}$=4600K, $\log
g$=1.2, $v_{\rm micro}$=2.6km s$^{-1}$, and [Fe/H]=$-3.1$.  
We used the equivalent widths obtained by this work and \citet{aoki09} 
for the abundance analysis.

The abundances of Fe {\small I} and Fe {\small II} (Ti {\small I} and
Ti {\small II}) derived with these atmospheric parameters are in good
agreement, confirming the adopted $\log g$ value. The iron abundance
of this star ([Fe/H] = $-$3.05) was derived from Fe {\small I} and Fe {\small II} lines.  We here adopt the \citet{asplund09} solar abundances.

\subsection{Neutron-capture elements}

The two resonance lines of Ba {\small II} at 4554~{\AA} and 4934~{\AA}
were measured by the present analysis, in addition to the two lines in
the red region (5853~{\AA} and 6141~{\AA}) measured by \citet{aoki09}.
The effects of a hyperfine splitting and isotope shifts are significant in
the abundance analysis of Ba, particularly for the strong resonance
lines.  We calculated the Ba abundances, assuming Ba isotope ratios
formed by the r-process and s-process, which were estimated from
solar-system material \citep{mcwilliam98,sneden08}.  We confirmed that
the abundances derived from the 4554~{\AA} and 4934~{\AA} assuming
r-process isotope ratios are significantly different (0.3 or 0.4 dex)
from those derived assuming the s-process one, while the results
derived from 5853~{\AA} and 6141~{\AA} are insensitive to the assumed
isotope ratios.  Therefore, we decided to adopt the Ba abundance
determined using the 5853 and 6141 lines, and to use the resonance
lines for comparisons.  
The final result was determined by assuming the
Ba isotope ratios for the s-process case, based on the abundance
pattern derived for this object in the following analysis.  
The derived Ba abundance, given in
table 2 ([Ba/Fe]), agrees well with that of \citet{aoki09}.

Among lighter neutron-capture elements, we determine the Sr abundance
([Sr/Fe] $=-1.4$) from the Sr {\small II} 4077~{\AA} line.  The
under-abundance of Sr is significant given the over-abundance of Ba.
Even if a large error in the equivalent width measurement (20~m{\AA})
for the Sr line is taken into account, the Sr abundance is at most
[Sr/Fe]$=-1.1$. Hence, the result that Sr is underabundant in this
object is robust.

Only upper limits of the abundances were derived for other
neutron-capture elements (Y, Zr, La, Ce, Nd, Eu and Dy) using the
error estimate for equivalent-width measurements for weak lines
provided by Norris et al. (2001) for a given resolution, S/N ratio and
sampling. The results were checked by the spectrum-synthesis technique.

Among the neutron-capture elements, the upper limit of the Eu abundance is
particularly important to constrain the origin of heavy elements in
this object. The upper limit of the Eu abundance [Eu/Fe]$<0.7$ was obtained
from the above estimate. However, the spectrum synthesis for the Eu
{\small II} line at 4129~{\AA} suggests that this is possibly an
underestimate (figure 2). We adopt the upper limit [Eu/Fe]$<0.9$ as the final
result (table 2).

This results in the upper limit of [Eu/Ba] to be $=+0.5$, which is
lower than the value expected for the r-process abundance ratio
([Eu/Ba]$_{\rm r-process}=+0.8$: e.g. Burris et al. 2000).  
The result is rather consistent with the s-process abundance ratio ([Eu/Ba]$_{\rm s-process}=-1.0$), 
although the constraint is not strong (figure 3).

Hence, the conclusion derived from the upper limit of Eu/Ba is that
the Ba excess of S~15--19 is not fully explained by the r-process, but
is at least partially contributed by the s-process. The low Sr/Ba
ratio ([Sr/Ba]$=-1.9$) also supports the significant contribution of
the s-process, because a low Sr/Ba ratio is expected from the s-process
(in an AGB star) at low metallicity, as a result of very high ratios
of neutrons to seed nuclei \citep{gallino98}, while the [Sr/Ba] value found for
r-process-enhanced EMP stars is $-0.4$ (e.g., CS~22892--052 : Sneden et al. 2003). 
This suggests that the s-process is the dominant contributor to 
the heavy neutron-capture elements in S~15--19.

\subsection{Other elements}

Another support for the s-process contribution is found in the carbon
abundance of this object. The carbon abundance was determined by a spectrum synthesis of the CH band at 4324~{\AA} (figure 4), adopting
the CH line data from \citet{aoki02}. The result clearly shows that
this star is carbon-enhanced ([C/Fe] = +1.0), which is a signature of
a contribution by AGB nucleosynthesis.

The carbon-isotope ratio was estimated from the CH lines at around 4200
{\AA}, as done by \citet{honda04}. The derived $^{12}$C/$^{13}$C ratio
is 4 $\pm$ 1, which is a typical value found in highly evolved red
giants (e.g., Charbonel 1995, Spite et al. 2006).
Such a low value is also found in some carbon-enhanced
metal-poor stars that have low temperature and low gravity (e.g., Ryan
et al. 2005).

We also estimate the nitrogen abundance from the CN band at
4215~{\AA}. Although the result is quite uncertain due to the low S/N
ratio at this wavelength, a large over-abundance of nitrogen is suggested
([N/Fe]$=+1.5\pm0.5$).

\subsection{Uncertainties}

We estimate the random errors in the analysis, which included the
uncertainties of adopted $gf$ values and the equivalent-width
measurements, as well as errors due to the uncertainties of stellar
parameters adopted in the analysis.

Random errors were estimated from the mean of the standard
deviations (1 $\sigma$) of the abundances derived from individual
lines for elements that had three or more lines available (Mg~{\small
  I}, Ca~{\small I}, Ti~{\small II}, Fe~{\small I}, and Fe~{\small
  II}).  We apply this value (0.19~dex) to the species for which only
one or two lines have been detected. We adopt 0.3 and 0.5~dex errors
for carbon and nitrogen abundances from the CH and CN bands from the
estimate by spectrum synthesis.

To estimate the errors due to the uncertainties of stellar parameters,
we examine the abundance analysis by changing the parameters, as given in
table 3. 
The change of C abundance due to the $T_{\rm eff}$ change was included
in the estimate of the effect on the N abundance determination from
the CN band.
These errors are not generally very significant, compared
with the random errors estimated above.

\section{Discussion}

Our chemical-abundance measurements for the red giant S~15--19 in the
Sextans dwarf spheroidal revealed that this star has excesses of
carbon ([C/Fe]$=+1$) and Ba ([Ba/Fe]=$+0.4$) but shows no clear excess
of r-process elements ([Eu/Ba]$<+0.5$). 
A radial velocity variation was
found between the two epochs of the past observing.  This indicates
that the excess of Ba in this star can be attributed to the s-process
in an AGB star that was a companion (primary) of the binary to which
S~15--19 belongs; also, carbon-rich material has been transfered from the
AGB star across the binary system.  The low Sr/Ba ratio
([Sr/Ba]$=-1.9$) also suggests the s-process at very low metallicity,
which efficiently yields heavier neutron-capture elements due to high
neutron-to-seed nuclei ratios \citep{gallino98}.  Hence, we conclude
that this object is a CEMP star with excesses of s-process elements
(CEMP-s: Beers \& Christlieb 2005).

Figure 5 compares the carbon and Ba abundances of this object with
those of Milky Way field stars, where carbon-enhanced stars are
emphasized \citep{aoki10}.  
A large number of CEMP stars in the field halo show very high carbon
and Ba abundances ([C/Fe]$>+2$, [Ba/Fe]$>+2$). 
The excesses of carbon and Ba of S~15--19 are not as significant as 
those of these stars. 
We note that such moderate excesses of carbon and
Ba are also found in CS~22892--052, which is the first example of
r-process-enhanced (r-II) stars in the field halo
\citep{sneden96}. However, the origin of neutron-capture elements of
S~15--19 is completely different from this object.

The relatively small excesses of carbon and Ba in S~15--19 suggest that 
the mass transfer from the AGB star is not as significant as the one 
that affected the CEMP stars with very high carbon abundances, or the 
carbon excess has been diluted during the evolution of S~15--19 along 
the red giant branch (see below). 

Among CEMP-s stars in the field halo, CS~30322--023 has similar
excesses of carbon and Ba to S~15--19. This object is a
carbon-enhanced ([C/Fe]$=+0.6$) extremely metal-poor ([Fe/H]$=-3.4$)
star with a moderate Ba excess ([Ba/Fe]$=+0.6$) studied by
\citet{masseron06} and \citet{aoki07}. Interesting features of this
object are a large enhancement of nitrogen ([N/Fe]$=+2.5$: Aoki et
al. 2007; +2.8: Masseron et al. 2006) and the low $^{12}$C/$^{13}$C (=4:
Masseron et al. 2006). This suggests that the carbon-enhanced material
has been affected by the CNO cycle. While \citet{masseron06} suggested
that this object could be a thermally pulsing AGB star, because of its
low temperature and low gravity, the carbon-enhancement might have
originated from the mass transfer from an AGB star, as in the case of
other CEMP-s stars, given the variations of radial velocities found 
during four years \citep{masseron06}. 
The low $^{12}$C/$^{13}$C ratio ($\sim 4$) and the high N abundance ([N/Fe]$\sim +1.5$) of S~15--19 suggest that this object has a similar origin
to CS~30322--023, although these results are rather uncertain. 
Whatever the evolutionary stage of S~15--19 and
CS~30322--023 is, the similarity suggests that natures of CEMP stars
could be common between the field halo and dwarf galaxies.

To derive a definitive conclusion for such discussions, the statistics of
carbon-enhanced stars in dwarf galaxies are required.
Unfortunately, however, studies of CEMP stars in dwarf galaxies are still quite limited. For Sextans, \citet{aoki09} reported a discovery of the CEMP star S~11--36
([Fe/H]$=-2.9$, [C/Fe]$=+1.9$) with moderate excess of Ba ([Ba/Fe]$=+0.8$). 
Although their data quality is insufficient to
discuss the details of the chemical composition of this object, the
existence of such a CEMP star supports the similarity of the stellar
population with carbon-excesses between the dwarf galaxy and the field halo. 

At higher metallicity, another carbon-enhanced star with large
excesses of neutron-capture elements was found in Ursa Minor dwarf
spheroidal (the star K: [Fe/H]$=-2.17$, [Ba/Fe]$=+1.37$,
[Eu/Ba]$=-0.33$ with strong C$_{2}$ bands) by \citet{shetrone01}.  At
lower metallicity, a carbon-enhanced star with no excess of heavy
elements was recently discovered in the UFdG SEGUE-1 (SEGUE1-7:
[Fe/H]$= -3.52$, [C/Fe]$=+2.3$, [Ba/Fe]$<-1.0$) by \citet{norris10}.
Such stars (so-called CEMP-no stars) are also found at extremely low
metallicity in the field halo. Even though the origins of CEMP-no
stars and CEMP-s stars could be quite different, both populations
also seem to appear in dwarf galaxies. Further study to investigate
the statistics of CEMP stars in dwarf galaxies, and comparisons with
that in the MW halo, will be useful to constrain the formation processes
of EMP stars.

S~15--19 is a unique object that shows a definitely large overabundance of Ba
among EMP stars in dwarf galaxies studied to date\footnote{The exception 
is S~11--36 that shows a large excess of Ba, though the abundance results 
are quite uncertain as mentioned in this section \citep{aoki09}.
This object also has large excess of carbon, suggesting the origin of
Ba is also the s-process.}.  
Since the origin
of Ba of this object turned out to be the s-process, no r-process enhanced
EMP star is known at present in dwarf galaxies.  This possibly
indicates that dwarf galaxy may have a history of the chemical evolution
different from MW. However, since the sample size of EMP stars in
dwarf galaxies studied with high-resolution spectroscopy is still
small, we cannot derive any definitive conclusion. Indeed, the
statistics for halo field stars reveals that the fraction of r-process
enhanced stars (r-II stars; [Eu/Fe]$>+1$) is on the order of 5\%
\citep{barklem05}, while Ba abundances have been measured for about 20
stars with [Fe/H]$<-2.5$ in dwarf galaxies, including UFdG
\citep{frebel10c}. Further searches for EMP stars and detailed
abundance measurements for dwarf galaxies are strongly desired in order to
clarify the chemical evolutions of dwarf galaxies and nucleosynthesis of
the r-process.

\bigskip
%Acknowledgement should be placed at end of main text.(NOT after the Appendix.)
We would like to thank Dr. A. Tajitsu for him help to our observation.
This work was supported by a Grant-in-Aid for Science Research from
MEXT and JSPS (grant 21740148, 18104003, and 19540240).

\begin{longtable}{lclrrrccc}
  \caption{Equivalent Width Measrements}\label{tab:first}
  \hline              
      Species & wavelength({\AA}) & L.E.P.(eV) & $\log gf$ & W(m{\AA}) & $\log \epsilon$  \\
      \hline
\endfirsthead
  \hline
      Species & wavelength({\AA}) & L.E.P.(eV) & $\log gf$ & W(m{\AA}) & $\log \epsilon$  \\
\hline
\endhead
  \hline
\endfoot
  \hline
\footnotemark[$*$] Taken from Aoki et al. (2009).
\endlastfoot
  \hline
      Na I & 5889.95 & 0.00 &  0.12 & 153.6*  &  3.22 &  \\
      Na I & 5895.92 & 0.00 & --0.18 & 137.6*  &  3.28 &  \\
      Mg I & 4571.10 & 0.00 & --5.39 &  84.5  &  5.11 &   \\
      Mg I & 4702.99 & 4.33 & --0.38 &  60.3  &  5.04 &   \\
      Mg I & 5172.68 & 2.71 & --0.45 & 203.9  &  5.10 &   \\
      Mg I & 5183.60 & 2.72 & --0.24 & 224.3  &  5.08 &   \\
      Mg I & 5528.40 & 4.35 & --0.34 &  55.4*  &  4.90 &  \\
      Ca I & 4318.65 & 1.90 & --0.21 &  48.4  &  3.71 &   \\
      Ca I & 5588.75 & 2.53 &  0.20 &  45.0*  &  3.91 &  \\
      Ca I & 6102.72 & 1.88 & --0.79 &  27.1*  &  3.78 &  \\
      Ca I & 6122.22 & 1.89 & --0.31 &  40.7*  &  3.55 &  \\
      Ca I & 6162.17 & 1.90 & --0.09 &  73.8*  &  3.80 &  \\
      Ca I & 6462.57 & 2.52 &  0.31 &  32.5*  &  3.53 &  \\
      Sc II & 4246.82 & 0.31 &  0.24 & 156.1  &  0.59 &  \\
      Sc II & 4415.56 & 0.60 & --0.67 &  90.9  &  0.58 &  \\
      Ti I & 4533.24 & 0.85 &  0.53 &  26.2  &  1.81 &    \\
      Ti I & 4981.73 & 0.85 &  0.56 &  47.8*  &  2.11 &  \\
      Ti II & 4443.79 & 1.08 & --0.70 &  93.3  &  1.63 &  \\
      Ti II & 4450.48 & 1.08 & --1.45 &  62.9  &  1.99 &  \\
      Ti II & 4468.51 & 1.13 & --0.60 & 110.4  &  1.89 &  \\
      Ti II & 4501.27 & 1.12 & --0.75 & 104.2  &  1.92 &  \\
      Ti II & 4533.97 & 1.24 & --0.77 & 105.5  &  2.09 &  \\
      Ti II & 4563.76 & 1.22 & --0.96 &  77.4  &  1.53 &  \\
      Ti II & 4571.97 & 1.57 & --0.53 &  86.7  &  1.73 &  \\
      Ti II & 5188.68 & 1.58 & --1.21 &  64.8  &  2.24 &  \\
      Ti II & 5336.77 & 1.58 & --1.70 &  43.7  &  2.32 &  \\
      Cr I & 5206.04 & 0.94 &  0.02 &  74.6  &  2.31 &  \\
      Cr I & 5208.42 & 0.94 &  0.16 &  79.2  &  2.24 &  \\
      Fe I & 4071.74 & 1.61 & --0.02 & 129.9  &  3.93 &  \\
      Fe I & 4132.06 & 1.61 & --0.65 & 102.0  &  3.91 &  \\
      Fe I & 4143.87 & 1.56 & --0.51 & 136.6  &  4.46 &  \\
      Fe I & 4187.04 & 2.45 & --0.55 &  71.8  &  4.27 &  \\
      Fe I & 4202.03 & 1.49 & --0.71 & 107.8  &  3.92 &  \\
      Fe I & 4250.79 & 1.56 & --0.71 & 110.8  &  4.05 &  \\
      Fe I & 4260.47 & 2.40 &  0.08 &  87.0  &  3.82 &  \\
      Fe I & 4337.05 & 1.56 & --1.70 &  75.6  &  4.36 &  \\
      Fe I & 4383.54 & 1.49 &  0.20 & 168.0  &  4.17 &  \\
      Fe I & 4442.34 & 2.20 & --1.26 &  64.4  &  4.51 &  \\
      Fe I & 4447.72 & 2.22 & --1.34 &  65.5  &  4.64 &  \\
      Fe I & 4459.12 & 2.18 & --1.28 &  58.5  &  4.42 &  \\
      Fe I & 4476.02 & 2.85 & --0.82 &  44.5  &  4.56 &  \\
      Fe I & 4494.56 & 2.20 & --1.14 &  67.0  &  4.42 &  \\
      Fe I & 4528.61 & 2.18 & --0.82 &  85.2  &  4.36 &  \\
      Fe I & 4531.15 & 1.49 & --2.16 &  64.1  &  4.52 &  \\
      Fe I & 4871.32 & 2.87 & --0.36 &  58.7  &  4.29 &  \\
      Fe I & 4872.14 & 2.88 & --0.57 &  46.2  &  4.34 &  \\
      Fe I & 4890.75 & 2.88 & --0.39 &  66.0  &  4.44 &  \\
      Fe I & 4891.49 & 2.85 & --0.11 &  75.2  &  4.27 &  \\
      Fe I & 4903.31 & 2.88 & --0.93 &  28.2  &  4.39 &  \\
      Fe I & 4918.99 & 2.85 & --0.34 &  56.0  &  4.21 &  \\
      Fe I & 4920.50 & 2.83 &  0.07 &  87.8  &  4.25 &  \\
      Fe I & 4939.69 & 0.86 & --3.25 &  69.7  &  4.86 &  \\
      Fe I & 4957.60 & 2.81 &  0.23 &  93.4  &  4.15 &  \\
      Fe I & 4994.13 & 0.92 & --3.08 &  65.6  &  4.70 &  \\
      Fe I & 5006.12 & 2.83 & --0.64 &  51.0  &  4.41 &  \\
      Fe I & 5012.07 & 0.86 & --2.64 & 106.7  &  4.82 &  \\
      Fe I & 5041.76 & 1.49 & --2.20 &  53.2  &  4.35 &  \\
      Fe I & 5049.82 & 2.28 & --1.42 &  41.4*  &  4.37 &  \\
      Fe I & 5079.74 & 0.99 & --3.22 &  58.1  &  4.82 &  \\
      Fe I & 5083.34 & 0.96 & --2.96 &  63.3  &  4.59 &  \\
      Fe I & 5123.72 & 1.01 & --3.07 &  37.6  &  4.39 &  \\
      Fe I & 5142.93 & 0.96 & --3.08 &  51.2  &  4.54 &  \\
      Fe I & 5150.84 & 0.99 & --3.07 &  45.9*  &  4.49 &  \\
      Fe I & 5151.91 & 1.01 & --3.32 &  41.3*  &  4.69 &  \\
      Fe I & 5166.28 & 0.00 & --4.20 &  69.0  &  4.70 &  \\
      Fe I & 5171.60 & 1.49 & --1.79 &  99.9  &  4.61 &  \\
      Fe I & 5194.94 & 1.56 & --2.09 &  53.9  &  4.32 &  \\
      Fe I & 5216.27 & 1.61 & --2.15 &  57.2  &  4.49 &  \\
      Fe I & 5232.94 & 2.94 & --0.06 &  63.7  &  4.12 &  \\
      Fe I & 5250.65 & 2.20 & --2.05 &  28.5*  &  4.66 &  \\
      Fe I & 5266.56 & 3.00 & --0.39 &  36.3  &  4.11 &  \\
      Fe I & 5269.54 & 0.86 & --1.32 & 158.6  &  4.43 &  \\
      Fe I & 5270.36 & 1.61 & --1.51 & 120.9  &  4.84 &  \\
      Fe I & 5324.18 & 3.21 & --0.24 &  43.1  &  4.32 &  \\
      Fe I & 5328.04 & 0.92 & --1.47 & 154.4  &  4.55 &  \\
      Fe I & 5328.53 & 1.56 & --1.85 &  78.8*  &  4.41 &   \\
      Fe I & 5339.93 & 3.27 & --0.68 &  26.0*  &  4.53 &  \\
      Fe I & 5341.02 & 1.61 & --2.06 &  78.0  &  4.67 &  \\
      Fe I & 5371.49 & 0.96 & --1.64 & 127.6  &  4.25 &  \\
      Fe I & 5397.13 & 0.92 & --1.99 & 112.1  &  4.26 &  \\
      Fe I & 5405.77 & 0.99 & --1.84 & 121.8  &  4.38 &  \\
      Fe I & 5434.52 & 1.01 & --2.12 & 117.6  &  4.60 &  \\
      Fe I & 5446.92 & 0.99 & --1.93 & 121.2  &  4.44 &  \\
      Fe I & 5455.61 & 1.01 & --2.09 & 125.2  &  4.70 &  \\
      Fe I & 5497.52 & 1.01 & --2.85 &  70.8  &  4.60 &  \\
      Fe I & 5501.46 & 0.96 & --2.95 &  58.2*  &  4.47 &   \\
      Fe I & 5506.78 & 0.99 & --2.80 &  82.3*  &  4.68 &   \\
      Fe I & 5569.62 & 3.42 & --0.54 &  29.5*  &  4.63 &   \\
      Fe I & 5615.64 & 3.33 & --0.14 &  55.0*  &  4.52 &   \\
      Fe I & 6136.62 & 2.45 & --1.40 &  37.9*  &  4.41 &   \\
      Fe I & 6137.69 & 2.59 & --1.40 &  20.3*  &  4.24 &   \\
      Fe I & 6191.56 & 2.43 & --1.60 &  35.1*  &  4.54 &   \\
      Fe I & 6430.85 & 2.18 & --2.01 &  31.0*  &  4.55 &   \\
      Fe II & 4508.29 & 2.86 & --2.31 &  39.5  &  4.52 &  \\
      Fe II & 4515.34 & 2.84 & --2.48 &  39.5  &  4.67 &  \\
      Fe II & 4583.84 & 2.81 & --1.74 &  59.4  &  4.19 &  \\
      Fe II & 4923.93 & 2.89 & --1.21 &  79.8  &  4.01 &  \\
      Fe II & 5018.43 & 2.89 & --1.23 & 101.1  &  4.35 &  \\
      Fe II & 5197.58 & 3.23 & --2.35 &  17.2  &  4.49 &  \\
      Fe II & 5234.63 & 3.22 & --2.15 &  14.5  &  4.19 &  \\
      Fe II & 5275.99 & 3.20 & --2.13 &  30.8  &  4.54 &  \\
      Fe II & 5316.62 & 3.15 & --2.02 &  33.4  &  4.42 &  \\
      Ni I & 5476.90 & 1.83 & --0.89 &  53.7  &  3.10 &   \\
      Zn I & 4722.15 & 4.03 & --0.39 &  14.6  &  2.10 &   \\
      Zn I & 4810.53 & 4.08 & --0.17 &  23.5  &  2.18 &   \\
      Sr II & 4077.72 & 0.00 &  0.17 & 105.3  & --1.56 &   \\
      Y II & 4883.69 & 1.08 &  0.07 & $<$37.4 & $<-$0.57 &   \\
      Zr II & 4317.32 & 0.71 & --1.38 & $<$46.5 & $<$1.10 &  \\
      Ba II & 4554.03 & 0.00 &  0.17 & 181.6  & --0.15 &  \\
      Ba II & 4934.08 & 0.00 & --0.15 & 173.6  & --0.19 &  \\
      Ba II & 5853.69 & 0.60 & --0.91 &  69.8*  & --0.41 &  \\
      Ba II & 6141.73 & 0.70 & --0.08 & 119.8*  & --0.44 &  \\
      La II & 4086.71 & 0.00 & --0.07 & $<$57.0 & $<-$1.09 &   \\
      Ce II & 4562.37 & 0.48 &  0.33 & $<$40.1 & $<-$0.49 &   \\
      Nd II & 4446.39 & 0.20 & --0.35 & $<$42.2 & $<-$0.38 &   \\
      Eu II & 4129.72 & 0.00 &  0.22 & $<$54.4 & $<-$1.84 &   \\
      Dy II & 4103.31 & 0.10 & --0.37 & $<$54.1 & $<-$0.51 &   \\

\end{longtable}

\begin{table}
  \begin{center}
  \caption{Chemical Abundances of S~15--19}\label{tab:first}
    \begin{tabular}{lrrlclllll}
      \hline
      Species & $\log \epsilon$ & [X/Fe] & N &  $\sigma$  \\
	  \hline
      C(CH) &  6.38    &   1.00   & --  & 0.3 \\
      N(CN) &  6.28    &   1.50   & --  & 0.5 \\
      Na I    &  3.25    &   0.13   & 2  & 0.19   \\
      Mg I   &  5.05    &   0.57   & 5  & 0.09 \\
      Ca I   &  3.71    &   0.45   & 6  & 0.15 \\
      Sc II   &  0.59    &   0.59   & 2  & 0.19   \\
      Ti I   &  1.96    &   0.11   & 2  & 0.19   \\
      Ti II  &  1.93    &   0.08   & 9  & 0.27 \\
      Cr I   &  2.27    & --0.32   & 2  & 0.19   \\
      Fe I   &  4.43    & --3.02   & 65 & 0.23 \\
      Fe II  &  4.37    & --3.08   & 9  & 0.21 \\
      Ni I   &  3.10    & --0.08   & 1  & 0.19   \\
      Zn I   &  2.14    &   0.59   & 2  & 0.19   \\
      Sr II   & --1.56   & --1.43   & 1  & 0.19   \\
      Y  II   & $<-$0.57 & $<$0.27 & 1  &  --    \\
      Zr II   & $<$1.10 & $<$1.57 & 1  &  --  \\
      Ba II   & --0.43   &  0.44   & 2  & 0.19 \\
      La II   & $<-$1.09 & $<$0.86 & 1  & --   \\
      Ce II   & $<-$0.49 & $<$0.98 & 1  & --  \\
      Nd II   & $<-$0.38 & $<$1.25 & 1  & --  \\
      Eu II   & $<-$1.63* & $<$0.90 & 1  & --   \\
      Dy II   & $<-$0.51 & $<$1.44 & 1  & --  \\
           \hline
   \multicolumn{4}{@{}l@{}}{\hbox to 0pt{\parbox{85mm}{\footnotesize
       Notes: N indicates the number of lines measured for the determination of abundance, and 
       $\sigma$ indicates the random error (see text).
       The upper limit of Eu is determined by spectrum synthesis. [Fe/H] values are given in the column of [X/Fe] for Fe{\small I} and Fe{\small II}.
       \par\noindent
     }\hss}}

    \end{tabular}
  \end{center}
\end{table}

\begin{table}
  \caption{Abundances changes from changing stellar parameters}\label{tab:error}
  \begin{center}
    \begin{tabular}{lrrrrrlllll}
      \hline
      Species & $\Delta$ $T_{\rm eff}$ & $\Delta$$\log g$ & $\Delta$[Fe/H] & $\Delta$$v_{\rm micro}$ & Total \\
%	  \hline
	       & (+150K)     & (+0.3dex) & (+0.3 dex)& (+0.3km) &(dex)   \\
	\hline
      C(CH)&   0.35   & --0.13    & +0.03  & --0.01   & 0.37 \\
      N(CN)&   0.19   & --0.10    & --0.07  &   0.00   & 0.21 \\
      Na I  &   0.15   & --0.04    & --0.03  & --0.14   & 0.21 \\
      Mg I  &   0.15   & --0.07    & --0.01  & --0.07   & 0.18 \\
      Ca I  &   0.10   & --0.02    & --0.01  & --0.02   & 0.11 \\
      Sc II &   0.10   &   0.06    & --0.02  & --0.15   & 0.19 \\
      Ti I  &   0.17   & --0.02    &   0.00  & --0.02   & 0.17 \\
      Ti II &   0.06   &   0.08    &   0.00  & --0.07   & 0.13 \\
      Cr I  &   0.17   & --0.03    & --0.01  & --0.05   & 0.18 \\
      Fe I  &   0.17   & --0.03    & --0.01  & --0.07   & 0.19 \\
      Fe II &   0.00   &   0.10    &   0.00  & --0.03   & 0.10 \\
      Ni I  &   0.17   & --0.02    & --0.01  & --0.03   & 0.17 \\
      Zn I  &   0.06   &   0.05    &   0.00  & --0.01   & 0.08 \\
      Sr II &   0.11   &   0.07    & --0.02  & --0.16   & 0.20 \\
      Y II  &   0.08   &   0.09    &   0.01  & --0.02   & 0.12 \\
      Zr II &   0.09   &   0.08    &   0.01  & --0.03   & 0.13 \\
      Ba II &   0.12   &   0.08    & --0.01  & --0.09   & 0.17 \\
      La II &   0.12   &   0.08    &   0.00  & --0.04   & 0.15 \\
      Ce II &   0.12   &   0.09    &   0.01  & --0.02   & 0.15 \\
      Nd II &   0.12   &   0.08    &   0.01  & --0.02   & 0.15 \\
      Eu II &   0.12   &   0.09    &   0.01  &   0.00   & 0.15 \\
      Dy II &   0.12   &   0.08    &   0.01  & --0.04   & 0.15 \\

     \hline
        \multicolumn{4}{@{}l@{}}{\hbox to 0pt{\parbox{85mm}{\footnotesize
       Notes: The total error is obtained by adding four uncertainties in quadrature.
       \par\noindent
     }\hss}}
    \end{tabular}
  \end{center}
\end{table}

\begin{figure}
  \begin{center}
    \FigureFile(90mm,30mm){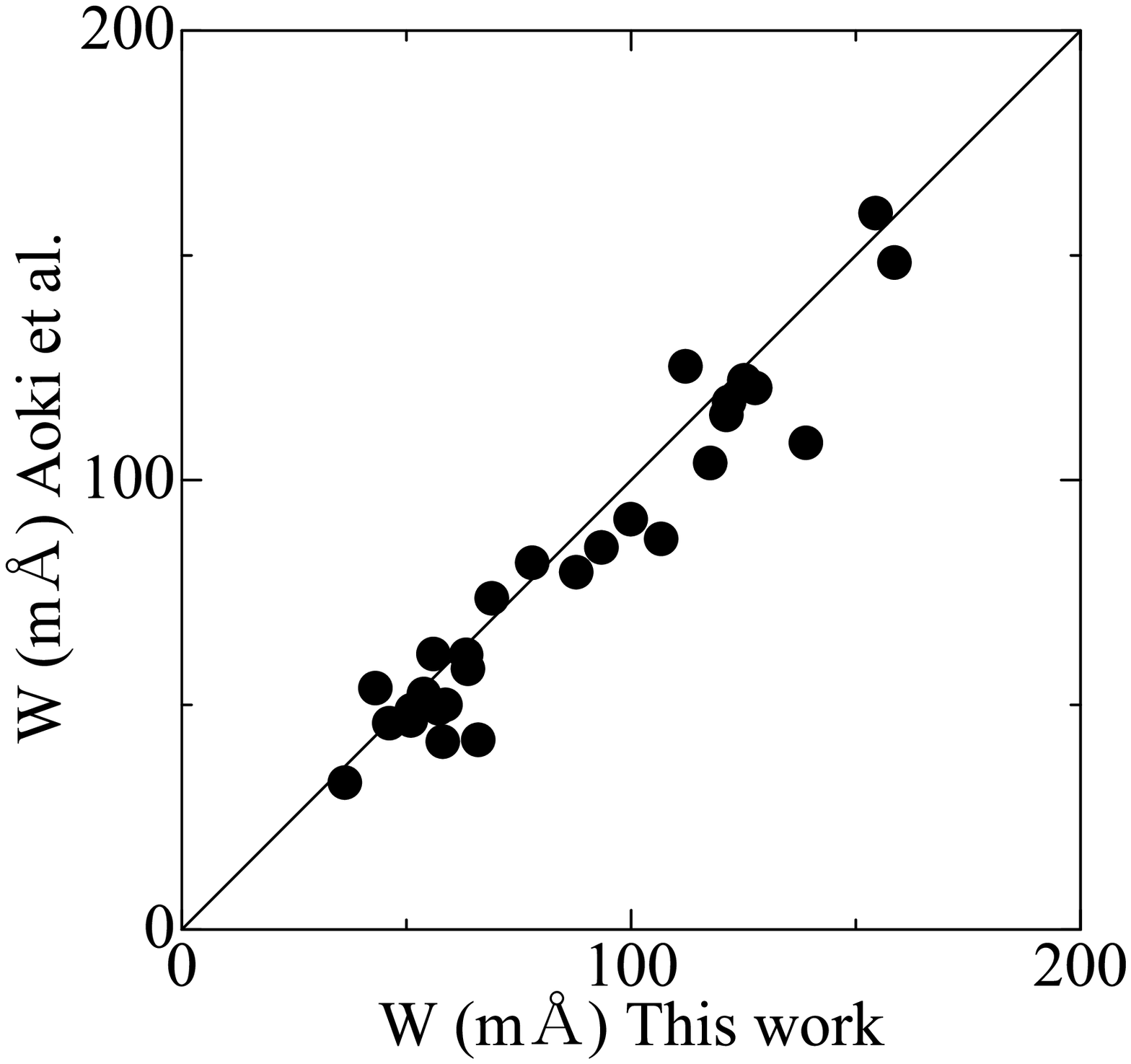}
  \end{center}
  \caption{Comparison of equivalent width (m{\AA}) measurements by Aoki et al. (2009) and this work.}\label{.....}
\end{figure}

\begin{figure}
  \begin{center}
    \FigureFile(90mm,30mm){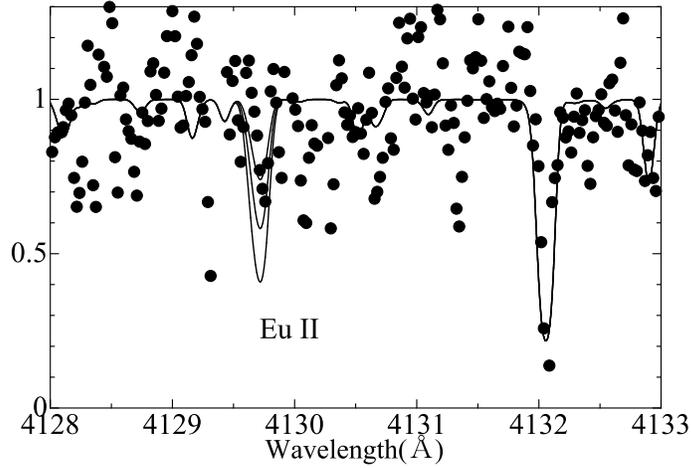}
  \end{center}
  \caption{Spectral region around the Eu feature at 4129 ~{\AA} for S~15--19. 
  The observed spectrum is shown by filled circles, and synthetic spectra with 
  three Eu abundances ([Eu/Fe] = +0.6, +0.9, +1.2) are shown by lines.
  The upper limit of Eu abundance ([Eu/Fe]$<+0.9$) is estimated from this
   comparison.}\label{.....}
\end{figure}

\begin{figure}
  \begin{center}
    \FigureFile(90mm,30mm){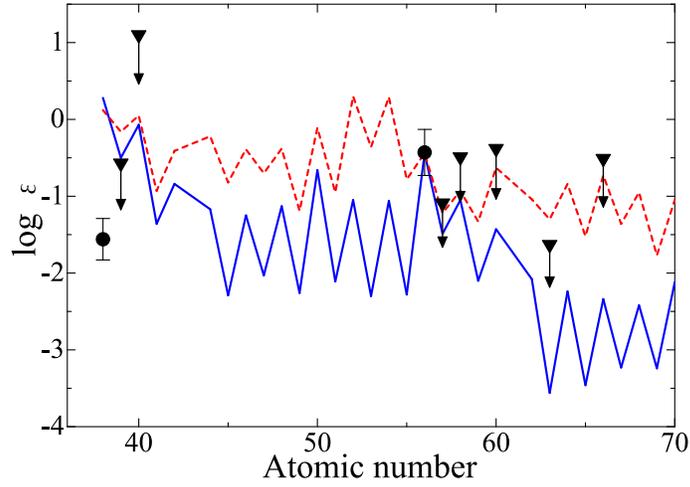}
  \end{center}
  \caption{Abundances of neutron-capture elements (filled circles), or upper limits (triangles), for S~15-19 as a function of the atomic number. The dashed and solid lines are the solar r- and s-process abundance patterns normalized at Ba.}\label{.....}
\end{figure}

\begin{figure}
  \begin{center}
    \FigureFile(90mm,30mm){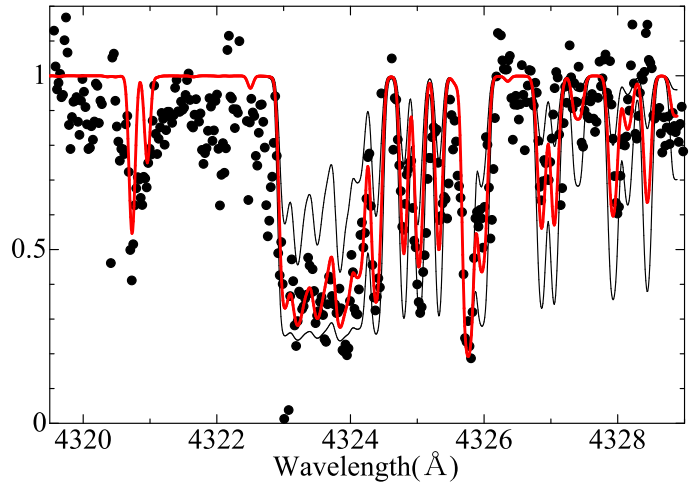}
  \end{center}
  \caption{Spectral region around the CH feature at 4323 ~{\AA} for S~15--19. 
  The observed spectrum is shown by filled circles, and synthetic spectra with 
  three carbon abundances ([C/Fe] = +0.5, +1.0, +1.5) are shown by lines.}\label{.....}
\end{figure}

\begin{figure}
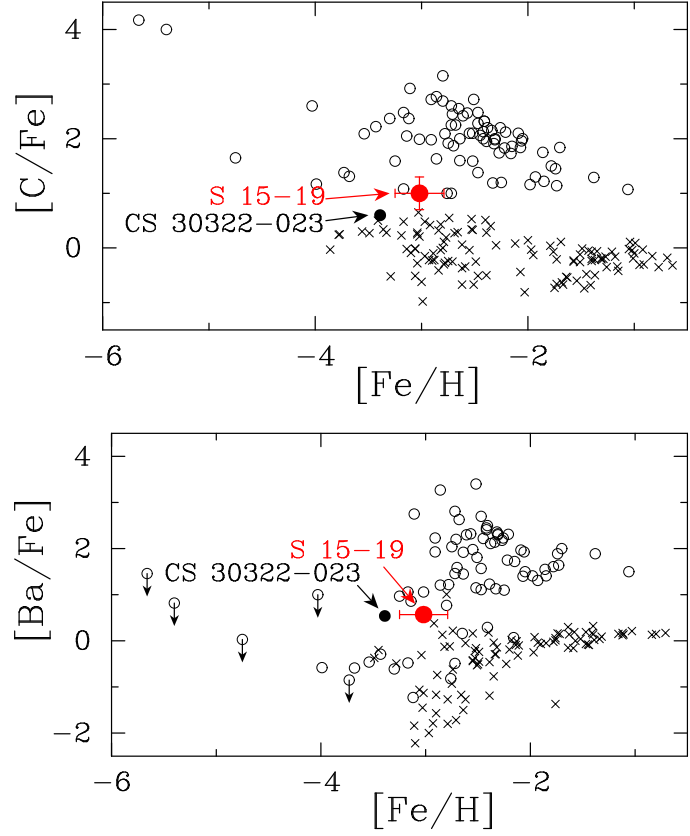

  \begin{center}
    \FigureFile(90mm,30mm){fig5a.ps}
    \FigureFile(90mm,30mm){fig5b.ps}
  \end{center}
  \caption{[C/Fe] and [Ba/Fe] as a function of [Fe/H]. Carbon-enhanced
    stars ([C/Fe]$\gtsim +1$) are shown by open circles, while
    non-carbon-enhanced stars are shown by crosses
    \citep{aoki10}. S~15--19 is shown by a large filled circle with
    error bars. CS~30322--023 (a filled circle) is classified into
    carbon-enhanced object, taking its evolutionary stage and the
    large nitrogen-enhancement into account (see text).}\label{.....}
\end{figure}

\end{document}